\documentclass[sigconf]{acmart}

\renewcommand\footnotetextcopyrightpermission[1]{} % removes footnote with conference information in first column
\newcommand\blfootnote[1]{%
  \begingroup
  \renewcommand\thefootnote{}\footnote{#1}%
  \addtocounter{footnote}{-1}%
  \endgroup
}

\usepackage[english]{babel}
\usepackage{blindtext}

\usepackage{lipsum}
\usepackage{booktabs}
\usepackage{paralist}
\PassOptionsToPackage{hyphens}{url}
\usepackage{hyperref}
\hypersetup{colorlinks, citecolor=blue, filecolor=blue, linkcolor=blue, urlcolor=blue}
\usepackage{multirow}
\usepackage{subfigure}
\usepackage{graphicx}

\usepackage[inline]{enumitem}

\newcommand{\afblock}[1]{\noindent{\textbf{#1}}}
\newcommand{\takeaway}[1]{\noindent{\textbf{Takeaway.}} \textit{#1}}

\usepackage{todonotes}

\begin{document}

\def\UrlBreaks{\do\/\do-}

\begin{abstract}
Booter services continue to provide popular DDoS-as-a-service platforms and enable anyone irrespective of their technical ability, to execute DDoS attacks with devastating impact. Since booters are a serious threat to Internet operations and can cause significant financial and reputational damage, they also draw the attention of law enforcement agencies and related counter activities. %
In this paper, we investigate booter-based DDoS attacks in the wild and the impact of an FBI takedown targeting 15 booter websites in December 2018 from the perspective of a major IXP and two ISPs. We study and compare attack properties of multiple booter services by launching Gbps-level attacks against our own infrastructure. To understand spatial and temporal trends of the DDoS traffic originating from booters we scrutinize $5$ months, worth of inter-domain traffic. 
We observe that the takedown only leads to a temporary reduction in attack traffic. Additionally, one booter was found to quickly continue operation by using a new domain for its website.

\end{abstract}

\title[DDoS Hide \& Seek: On the Effectiveness of a Booter Services Takedown]{DDoS Hide \& Seek: On the Effectiveness of a\\ Booter Services Takedown}

\author{Daniel Kopp}
\affiliation{%
  \institution{DE-CIX}}

\author{Matthias Wichtlhuber}
\affiliation{%
  \institution{DE-CIX}}

\author{Ingmar Poese}
\affiliation{%
  \institution{BENOCS}}

\author{Jair Santanna}  
\affiliation{%
  \institution{University of Twente}}

\author{Oliver Hohlfeld}  
\affiliation{%
  \institution{Brandenburg University of Technology}}

\author{Christoph Dietzel}
\affiliation{%
  \institution{DE-CIX / MPI for Informatics}}

\renewcommand{\shortauthors}{Daniel Kopp, Matthias Wichtlhuber, Ingmar Poese, Jair Santanna, Oliver Hohlfeld and Christoph Dietzel}

\maketitle

\blfootnote{Accepted for publication at ACM Internet Measurement Conference (IMC), 2019.}

\section{Introduction}\label{sec:intro}
\afblock{The DDoS threat.} Known for at least two decades~\cite{dittrich99ddos}, Distributed Denial-of-Service (DDoS) attacks have become a major security threat to the continuous operation of the Internet~\cite{akamaiDDoSreport,github,jonker17ddos}. Their goal is to disrupt services by consuming more critical resources than available, e.g., computing power or network bandwidth.
Beyond the web, modern DDoS attacks can overwhelm cloud services~\cite{sipgate14} or congest backbone peering links~\cite{OVH-Large-attack}.
The motivation for launching attacks ranges from 
financial~\cite{ransomware,holz12businessddos},
to political~\cite{bank-attack,russian-elections-2012}, 
cyber warfare~\cite{estonia2019misc,ukraine2019misc}, 
smoke screen for other attack types~\cite{KasperskyLab2019misc}, 
and even teenagers attacking their schools%
~\cite{santanna2015booters}.
To scale, DDoS amplification attacks~\cite{rossow14amplification,ryba15amplification} abuse protocol design (flaws)---e.g., NTP, DNS, SNMP, and Memcached~\cite{czyz2014,rossow14amplification,memcached-Akamai,morales17tbps18}---where a relatively small request can trigger a significantly larger response (up to $\times50\,000$). Spoofed source IP addresses~\cite{lichtblau17spoofing,IMC2009Spoofer,BeverlySRUTI2005,Moore2001} allow traffic to be reflected to the target~\cite{uscert18amplification}. Thus, attacks are increasing in size and sophistication~\cite{mckeay2017akamai}. A few years back, the largest reported attacks peaked just below $300$ Gbps~\cite{princesmallddos13,princeddos14}, whereas DDoS attacks have now reached the Tbps level~\cite{mirai-botnet-usenix-security,2016-Mirai-attack,morales17tbps18,github}.
\afblock{Booters as DDoS tool.} DDoS-as-a-service providers, also referred to as \emph{booters} or \emph{stressers}, provide a simple web interface and enable anyone to launch attacks~\cite{karami13bootersworkshop,karami2013booters} at a low cost of just \$0-\$5~\cite{santanaCharacterizing, insideBooters}. On the other end of the scale, booters also claim to \textit{offer} large-scale attacks for hundreds of USD~\cite{insideBooters}. %

\afblock{Booter-related work.} An extensive body of research on booters is available. These studies cover multiple lines of research including analyses of (1) the booters' leaked databases \cite{karami2013booters, twbooterLeet,insideBooters}, (2) booter attacks~\cite{zand2017demystifying,santanna2015booters,bukac2015service,wwwDDoSBusiness}, (3) victims of booters~\cite{noroozian2016gets}, (4) honeypots of servers commonly used for performing booter attacks~\cite{kramer2015raid,thomas2017ecrime}, (5) the usage of these honeypots for attribution purposes~\cite{krupp2017linking}, (6) TLS certificates used by booters~\cite{booterTLS}, (7) booter blacklists and their origins~\cite{booterBlacklist, booterListGeneration,booterWebsites,zhang2018booter}, (8) the usage of these blacklists to understand the booter market \cite{santanna2017quiet}, (9) ethical and legal aspects related to booters~\cite{hutchings2016exploring,douglas2017booters}, and (10) the impact of law enforcement operations on booters from a commercial perspective~\cite{booted,bootingBooters}.

\afblock{Our contribution.} In this first of its kind study, we shed light on an FBI led operation of 15 booter domains in December 2018~\cite{krebs-seizure, usdoj2018misc}. Instead of analyzing the effect of a financial intervention, we take an empirical perspective on the impact of this event on the DDoS attack traffic observed through the lens of three major networks: tier-1 and tier-2 ISPs, and a major IXP. 

Our study takes as its starting point the use of four popular booters to attack ourselves. This self-attack provides an up-to-date picture of booter capabilities and is the first study to investigate premium (VIP) services and their promises.
We derive characteristics of these self-attacks to further investigate tier-1 and tier-2 ISP and IXP network traffic for DDoS attacks.
We study spatial and temporal trends and present the existence and extent of DDoS attacks. This provides an overview of the current DDoS threat landscape. Finally, our study focuses on network traffic in the time frame of the FBI takedown, by applying the characteristics we learn from our self-attack approach and the investigation of network traffic of our vantage points.
In summary, our major contributions are:
\begin{itemize}[noitemsep,topsep=5pt,leftmargin=9pt]
\vspace{-0.5em}
\item We investigate the anatomy of a booter attack by launching attacks against our infrastructure. We were able to observe high attack traffic volumes of up to $20$ Gbps. 
\item 
We present an overview of the current Internet threat landscape through the lens of three major networks: tier-1 and tier-2 ISPs and a major IXP.
We observe constant DDoS attacks, at thousands of victims and traffic rates of up to $600$ Gbps.
\item We seize the unique opportunity to study the effectiveness of an FBI takedown targeting 15 booter services in December 2018. 
The takedown immediately reduced the DDoS amplification traffic to reflectors.
However, it did not have any significant effect on DDoS traffic hitting victims or on the number of attacks observed.
\end{itemize}

\section{Vantage Points}\label{sec:data}

Our study is based on three vantage points---a major IXP, a tier-1 ISP, and a tier-2 ISP---that provide a unique perspective on DDoS attack traffic in the wild. 
None of the data sets contain any payload. %
In addition, we perform active measurements of large sets of domains within an observatory~\cite{netrayPoster} to identify booter websites.
 
\afblock{Major IXP.} Anonymized and sampled IPFIX traces captured at a major Internet Exchange Point (IXP) between Oct. 27, 2018 and Jan. 31, 2019 with $834$\,B flows were made available to us. %

\afblock{Tier-1 ISP.}
We obtained Netflow traces from all border routers (ingress only) of a tier-1 ISP.
IP addresses are anonymized and filtered by protocol and port, resulting in $6.6$\,B flows records for the period of Dec. 12 to Dec. 30, 2018. The trace contains traffic to \emph{i)} fixed-line end-users, \emph{ii)} cellular customers, and \emph{iii)} transit traffic. Traffic from end-users and customers was not included.

\afblock{Tier-2 ISP.}
The second ISP dataset was anonymized and filtered in the same way as the tier-1 ISP.
However, ingress and egress traffic is available in this data, meaning that end-user and customer sourced traffic is included.
This results in $470$\,M flow records from Sept. 27, 2018 to Feb. 2, 2019.

\afblock{IXP Observatory.}
To study booter properties by performing self-attacks, we set up and operate an IXP-based DDoS observatory. %
It comprises a measurement AS operated by us that is connected to an IXP via a $10$GE link. The AS interconnection consists of multilateral peerings at the IXP and a transit link over the same physical interface. Data collection is performed directly at the IXP platform (sampled) and at the measurement AS itself (unsampled).

\afblock{DNS and HTTPS observatory.}
To study the rise and fall of booter websites, we use weekly crawls of all $\sim$$140$M .com/.net/.org domains by obtaining zone files and performing weekly DNS resolutions and HTTPS website snapshots during January 2018 until May 2019. The website snapshots enable us to identify booter websites.
\section{Booter: Victim's Perspective}\label{sec:booter}

We start by taking a victims' perspective to study the potential damage that booter-based DDoS attacks can (a) directly cause to their target (thereby updating earlier findings on booter attack characteristics~\cite{zand2017demystifying,santanna2015booters,bukac2015service,wwwDDoSBusiness}) and (b) the collateral damage to Internet infrastructure caused by carrying attack traffic.
We do so by purchasing services from popular booters to attack our dedicated measurement infrastructure at an IXP between April and September 2018.
This provides us with a unique picture of \emph{current} booter service capabilities in the wild: how much DDoS traffic they can generate in light of powerful Tbps-level attacks.
Our study provides the first look into how reliable the promises of these services are, e.g., premium membership benefits, promised attack protocols, and duration.
With our measurement infrastructure, we can draw conclusions about the DDoS traffic landscape.
Ultimately, we utilize the self-attack to identify attack characteristics to later discover DDoS attack traffic at our vantage points.

\subsection{Self-Attack Approach}

\newcommand{\serviceTested}{$\boldsymbol\checkmark$} %
\newcommand{\serviceAvailable}{$\color{gray}\checkmark$}
\newcommand{\seized}{$\checkmark$}

\begin{table}[t]
\centering
\small
\begin{tabular}{@{}cclcccccc@{}}
\multicolumn{1}{l}{\rotatebox{90}{Booter}} & \multicolumn{1}{l}{\rotatebox{90}{Seized}} & Time & \multicolumn{1}{l}{\rotatebox{90}{NTP}} & \multicolumn{1}{l}{\rotatebox{90}{DNS}} & \multicolumn{1}{l}{\rotatebox{90}{CLDAP}} & \multicolumn{1}{l}{\rotatebox{90}{mcache}} & \multicolumn{1}{l}{\rotatebox{90}{non-VIP}} & \multicolumn{1}{l}{\rotatebox{90}{VIP}} \\ \midrule
A & \seized & Apr, Aug & \serviceTested & \serviceAvailable & \serviceAvailable & \serviceAvailable & \textbf{\$8.00} & \$250 \\
B & \seized & Jun-Sep & \serviceTested & \serviceAvailable & \serviceTested & \serviceTested & \textbf{\$19.83} & \textbf{\$178.84} \\
C &  & Apr-May & \serviceTested & \serviceAvailable &  &  & \textbf{\$14.00} & \$89 \\
D &  & May & \serviceTested & \serviceAvailable &  &  & \textbf{\$19.99}  &  \$149.99 \\ \bottomrule
\end{tabular}
\caption{Booters used to attack our measurement AS. Booter services used for self-attack in Section~\ref{sec:booter} indicated in bold.}
\vspace{-3.05em}
\label{tab:booters}
\end{table}

\begin{figure*}[ht]
{
  \centering
  \subfigure[DDoS attacks by paid non-VIP services.]{
    \includegraphics[height=0.25\linewidth]{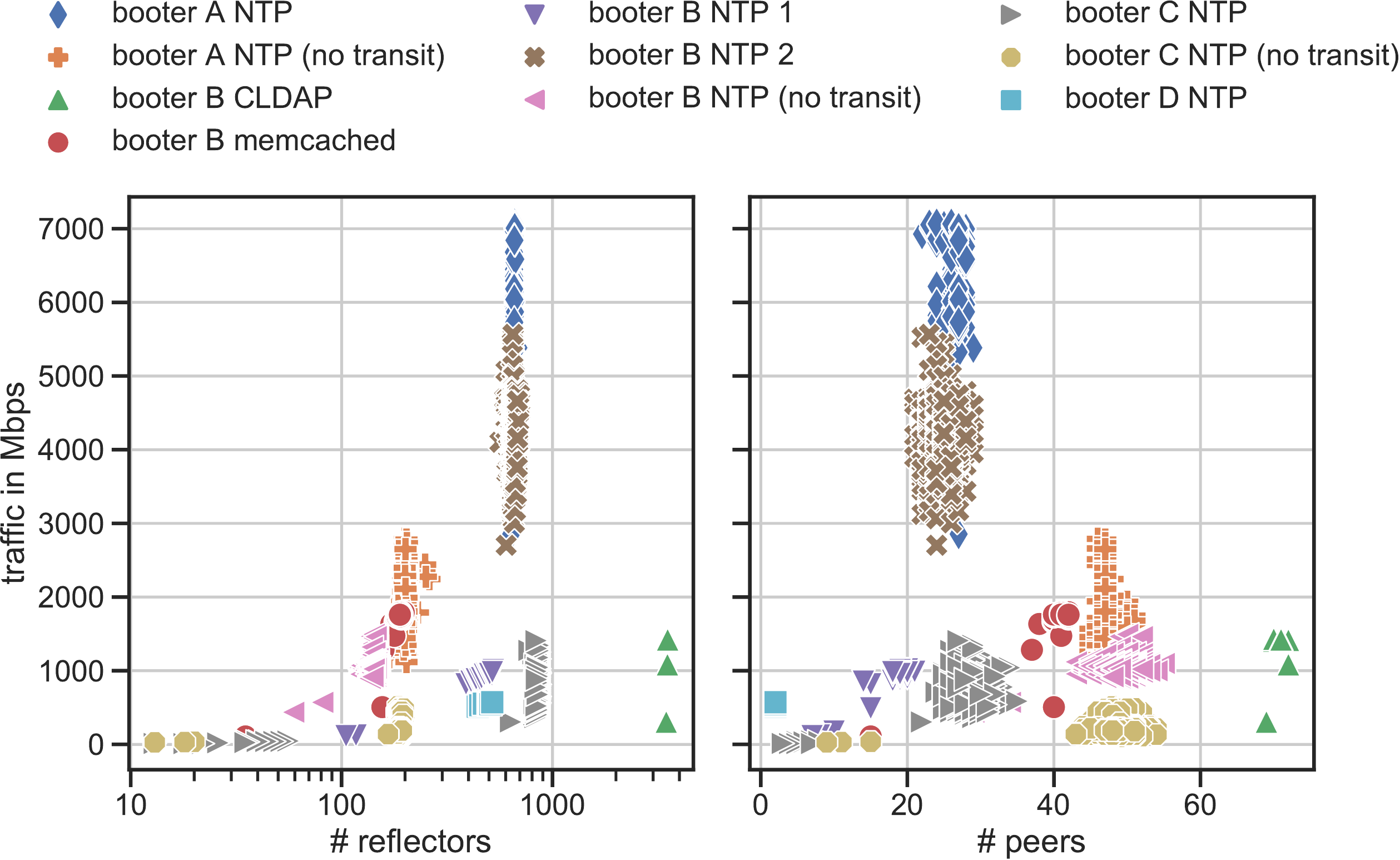}
    \label{fig:ddos_overview}
  }%
  \hfill
  \subfigure[Selected DDoS, measured at the IXP.]{
    \includegraphics[height=0.2\linewidth]{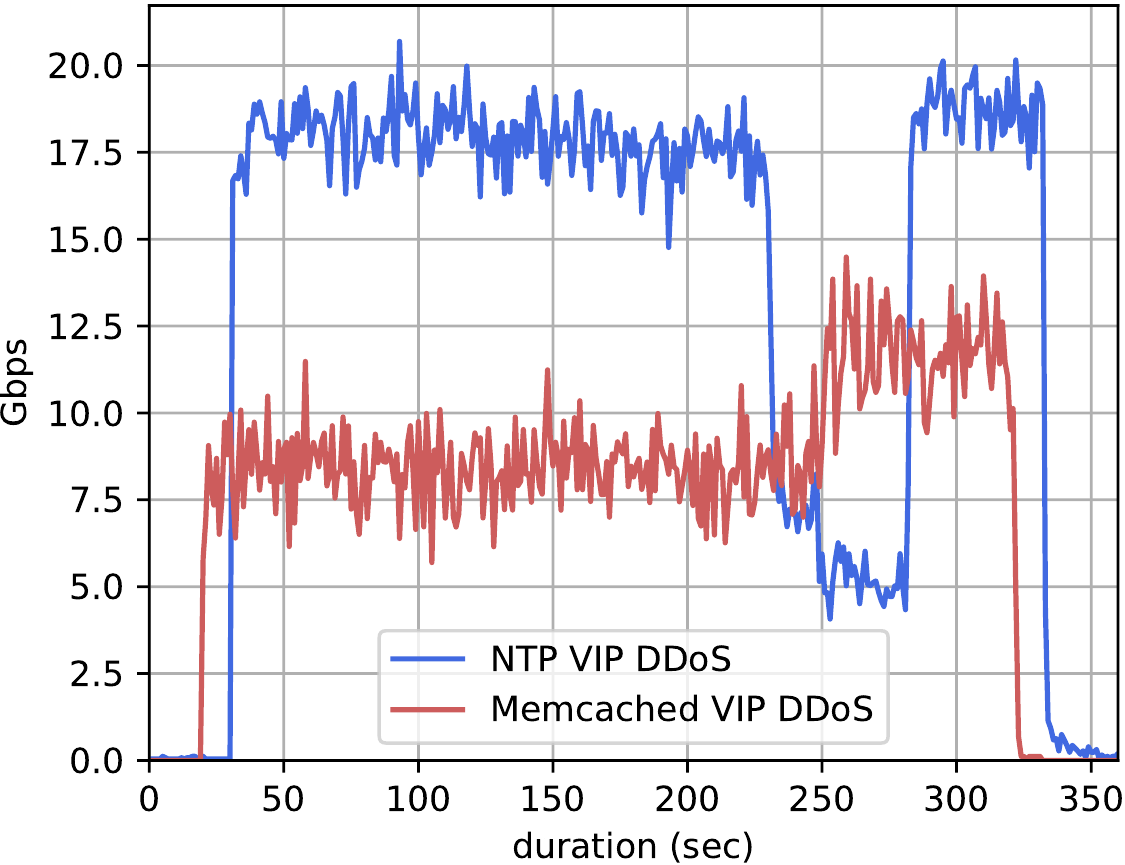}
    \label{fig:ddos_details}
  }
  \hfill
  \subfigure[Overlap of NTP reflectors over time.]{
    \includegraphics[height=0.25\linewidth]{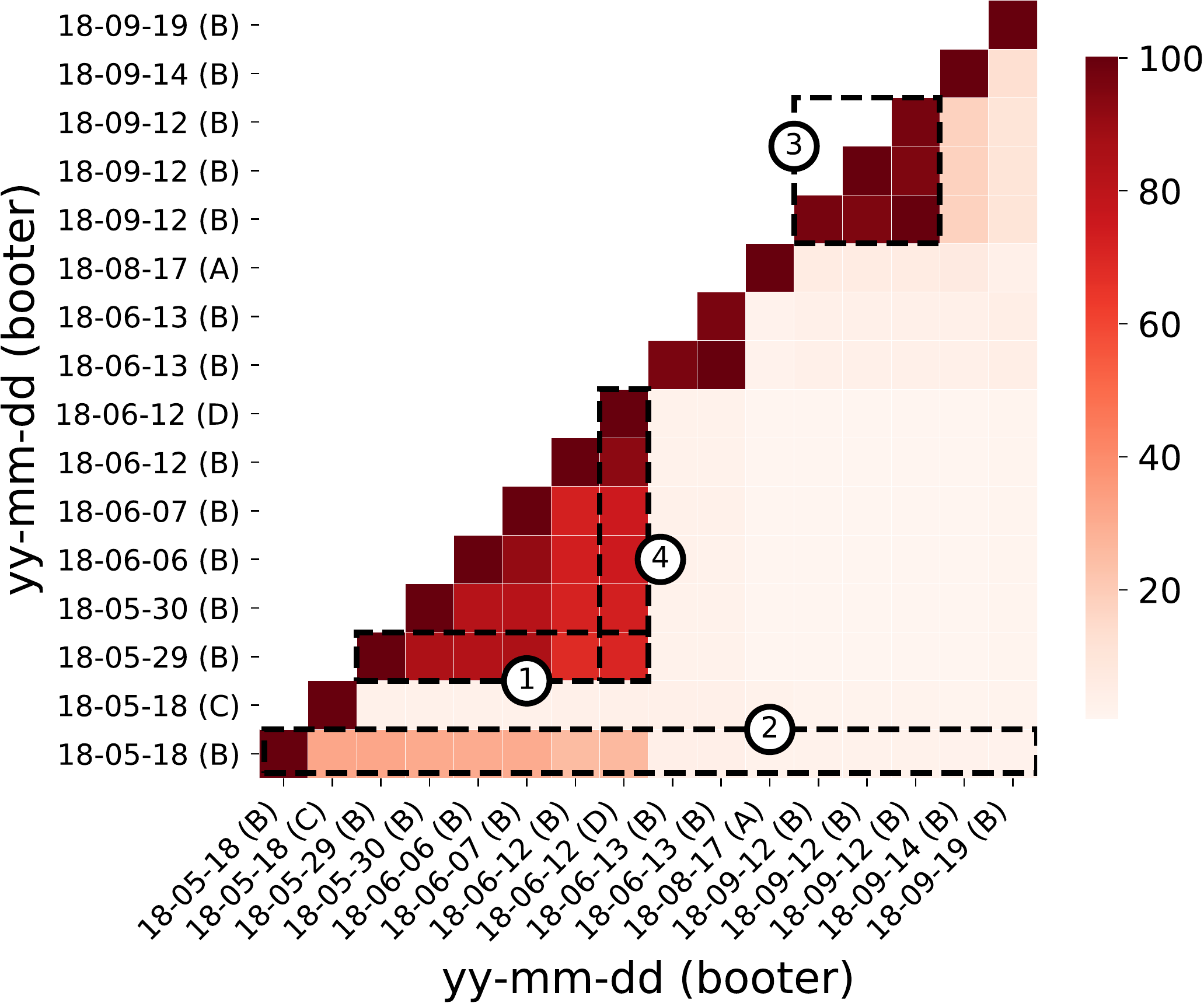}
    \label{fig:refl_hmap}
  }%
  \vspace{-1.5em}
  \caption{Measurements of Self-Attacks}
\vspace{-1em}
  \label{fig:eval}
}
\end{figure*}

\afblock{Selected booter services.}
We select $4$ popular booters (see Table~\ref{tab:booters}) from the booter blacklist~\cite{booterBlacklist} based on their Alexa website rank (booter names anonymized). Two of the selected booters ($A$ \& $B$) were later seized by the FBI-lead takedown. Three are still in operation (seized booter $A$ started using a new website after the takedown).
We purchase paid services from all booters including cheaper (non-VIP) and one more expensive premium package (VIP) from booter $B$. 
We use all booter services to launch attacks against our measurement infrastructure. However, for the remainder of this section we only consider attacks with relevant traffic volumes. These are mostly NTP based amplification attacks, whereas also CLDAP, DNS, and memcached-based attacks are offered.

\afblock{Attacking our infrastructure.}
For our analysis, we passively capture all traffic of the measurement platform. In addition, we obtain sampled flow traces of the IXP for traffic directed to our server and are therefore able to measure attack traffic exceeding the capacity of $10$ Gbps. The BGP router of our measurement platform announces a $/24$ IPv4 prefix and peers with a transit provider and all IXP customers in a multilateral peering configuration via the IXP's BGP reflector~\cite{IMC2014-RS}. This provides us with a similar network setup compared to small to medium-sized organizations connected to the Internet.
For each attack, we select a new IP out of our $/24$ prefix to isolate each individual measurement and to not confuse different attacks within our traffic captures.
We perform a post mortem analysis of the passively measured attacks and derive the attack traffic volume and their network properties (e.g., number of servers used for reflection, number of ASes handing over traffic).

\afblock{Ethical considerations.}
We perform controlled DDoS experiments towards our  measurement platform.
To comply with measurements ethics we 
\begin{inparaenum}[(a)]
\item inform and synchronize with
national authorities regarding legal/ethical implications of buying booter services,
\item minimize payments to booter services by limiting the number of different booters and offered service plans we buy (indicated in bold in Table~\ref{tab:booters}), 
\item inform and synchronize with the IXP operator and upstream provider about attacks, 
\item take precaution that sufficient IXP network bandwidth is available to minimize the influence on other IXP members, 
\item use an experimental AS with no customer traffic, 
\item utilize an unused $/24$ prefix that was allocated and announced only for the experiment, 
\item are prepared to shut down the experimental AS and immediately stop attack traffic by withdrawing and blackholing the $/24$ in case of unexpected high traffic volumes or IXP members being negatively effected by our experiment (never occurred), and 
\item minimize the experiment duration.
\end{inparaenum}

\subsection{Self-Attack Observations}\label{sec:self-attack}

\afblock{Attack traffic of non-VIP services.}
We show the results of $10$ self-attacks on our measurement platform by non-VIP booter services in Figure~\ref{fig:ddos_overview}. It shows the received traffic volume per second (y-axis) for each attack. On the x-axis we display the number of observed reflectors (left plot) together with the number of neighboring ASes from which we receive the traffic (right plot). Each data point represents one second of a measurement.
In terms of attack traffic volume, we find that during the attacks traffic levels of up to $\sim$$2000$ Mbps are prevalent with a mean of $1440$ Mbps but booters $B$ and $A$ peak at $7078$ Mbps. 
These are the highest traffic levels reported during a non-VIP booter attack to date~\cite{santanna2015booters}. 
When we focus on the number of reflectors and peers from which the traffic is coming, we find that most booter attacks utilize between $\sim$$100$ and $\sim$$1000$ reflectors distributed over $20$--$55$ peer ASes (avg. $346$ and $27$ respectively). 
However, when we instantiate booter $B$ to use the connectionless LDAP (CLDAP) protocol the number of reflectors is $3519$ distributed across $72$ peer ASes. 
Thus, we learn that the protocol used for amplification seems to have an effect on the number of reflectors and IXP members transmitting traffic. Finally, we observe that NTP amplification attacks are the most potent attacks delivered by the booters included in our tests.

We next study how the attack traffic is handed over to our AS at the IXP, i.e., which fraction is received via transit and via peering.
To first study the maximum traffic that can be received via peering at the present IXP, we perform three attacks solely via IXP peering and with \emph{disabled transit} link (indicated as ``no transit'') in Figure~\ref{fig:eval}).
This enforces the usage of peering links even if the transit link would have been a better routing option.
In this case, the number of individual IXP members (peers) sending traffic increases from below 30 to above 40 when the transit link is deactivated.
While the handover traffic spreads now over more peers, the absence of a full routing table limits the reachability of our AS when the transit link is disabled.
Consequently, we receive less attack traffic, e.g., the NTP attack volume of booter A decreases from up to 7 Gbps to less than 3 Gbps (see Figure~\ref{fig:ddos_overview}).
For NTP attacks with \emph{enabled transit}, we now receive most traffic through the transit link (avg. $80.81\%$) compared to the multilateral peerings at the IXP (avg. $19.19\%$).
Thus, the attack traffic volumes reported in Section~\ref{sec:ixp_ddos} captured at the peering platform of the IXP will likely underestimate the true attack sizes as the traces do not contain the customers' transit links.

\afblock{Attack traffic of VIP services.} Booter services advertise higher priced premium services. For booter B, VIP offerings charge 178.84\$  compared to 19.83\$ and promise higher attack traffic rates of $80$--$100$ Gbps instead of $8$--$12$ Gbps for non-VIP services. We validate this claim and launch \emph{two} VIP attacks from booter $B$. In Figure~\ref{fig:ddos_details}, we show that the NTP (blue line) and Memcached (red line) attacks generated traffic rates with a peak of about $20$ Gbps and $10$ Gbps, respectively. We configure both attacks to last for $5$ min. The sudden drop in attack for the NTP traffic is due to a flapping BGP session with our transit provider because of the saturation of our measurement interface.
\begin{figure*}[ht]
{
  \centering
  \subfigure[CDF/PDF of NTP pkt. sizes in IXP data.]{
    \includegraphics[height=0.28\linewidth]{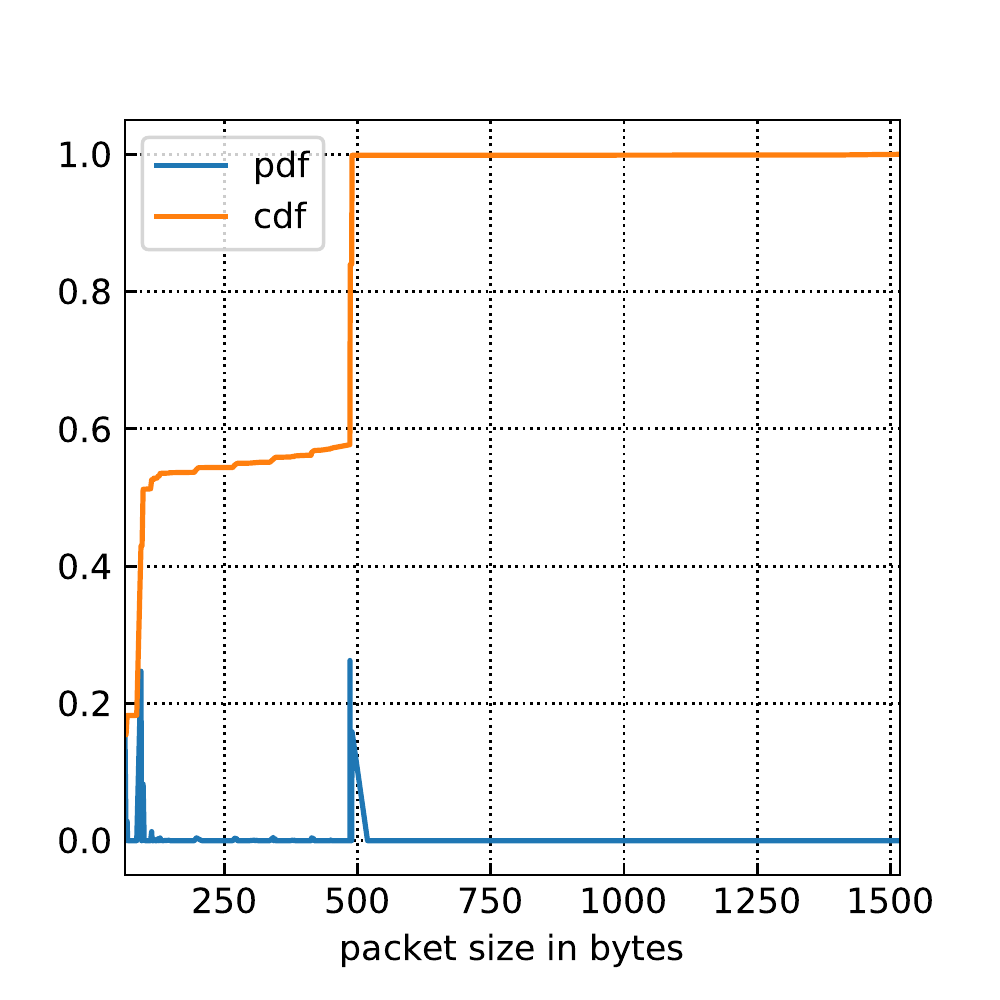}
    \label{fig:NTP_packet_size_cdf}
  }
  \hfill
  \subfigure[Traffic and reflectors per dest. IP at ISPs/IXP.]{
    \includegraphics[height=0.25\linewidth]{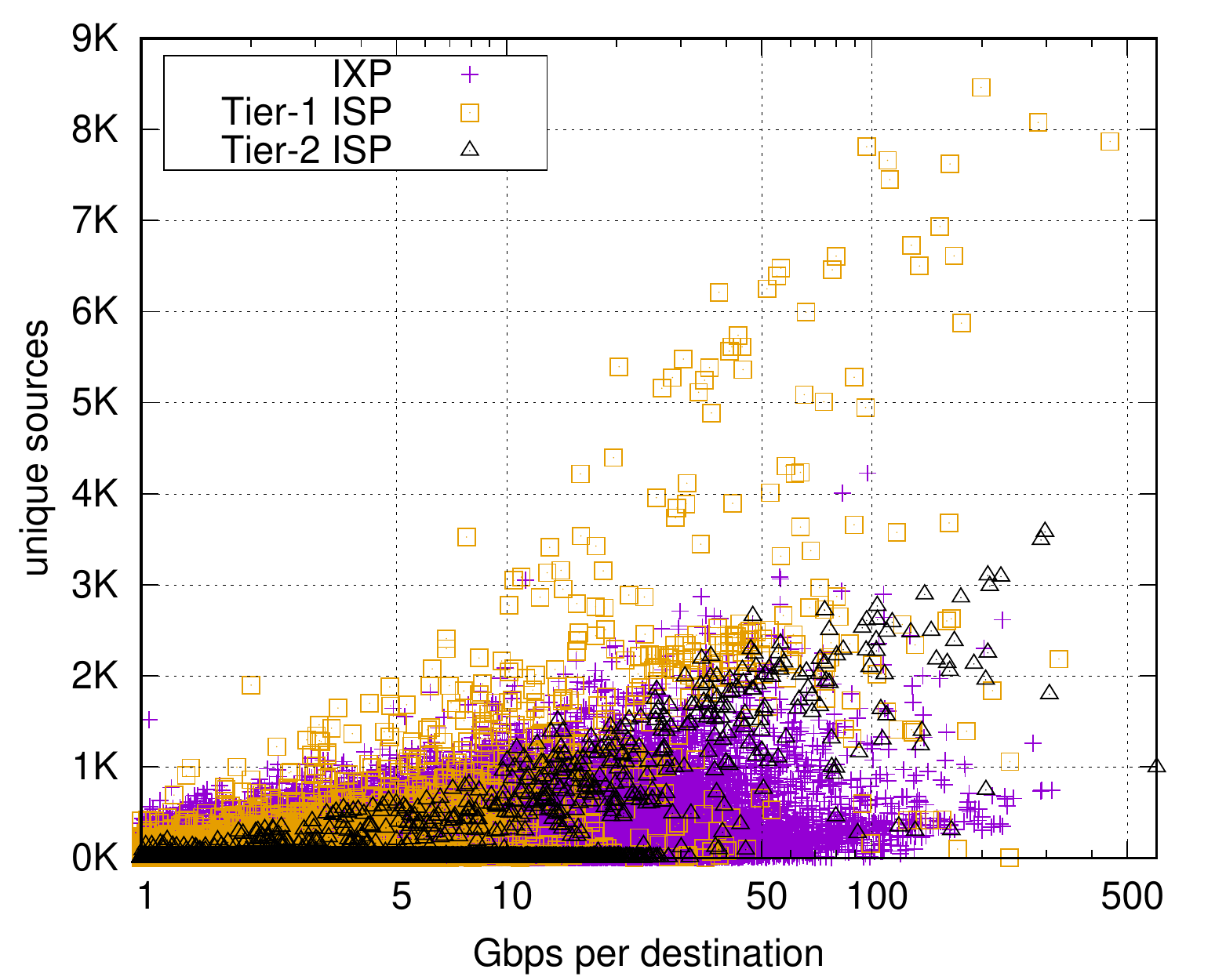}
    \label{fig:ISP_AttackSeverity_scatterplor}
  }
  \hfill
  \subfigure[CDF no. of reflectors per dest. IP at ISPs/IXP.]{
    \includegraphics[height=0.28\linewidth]{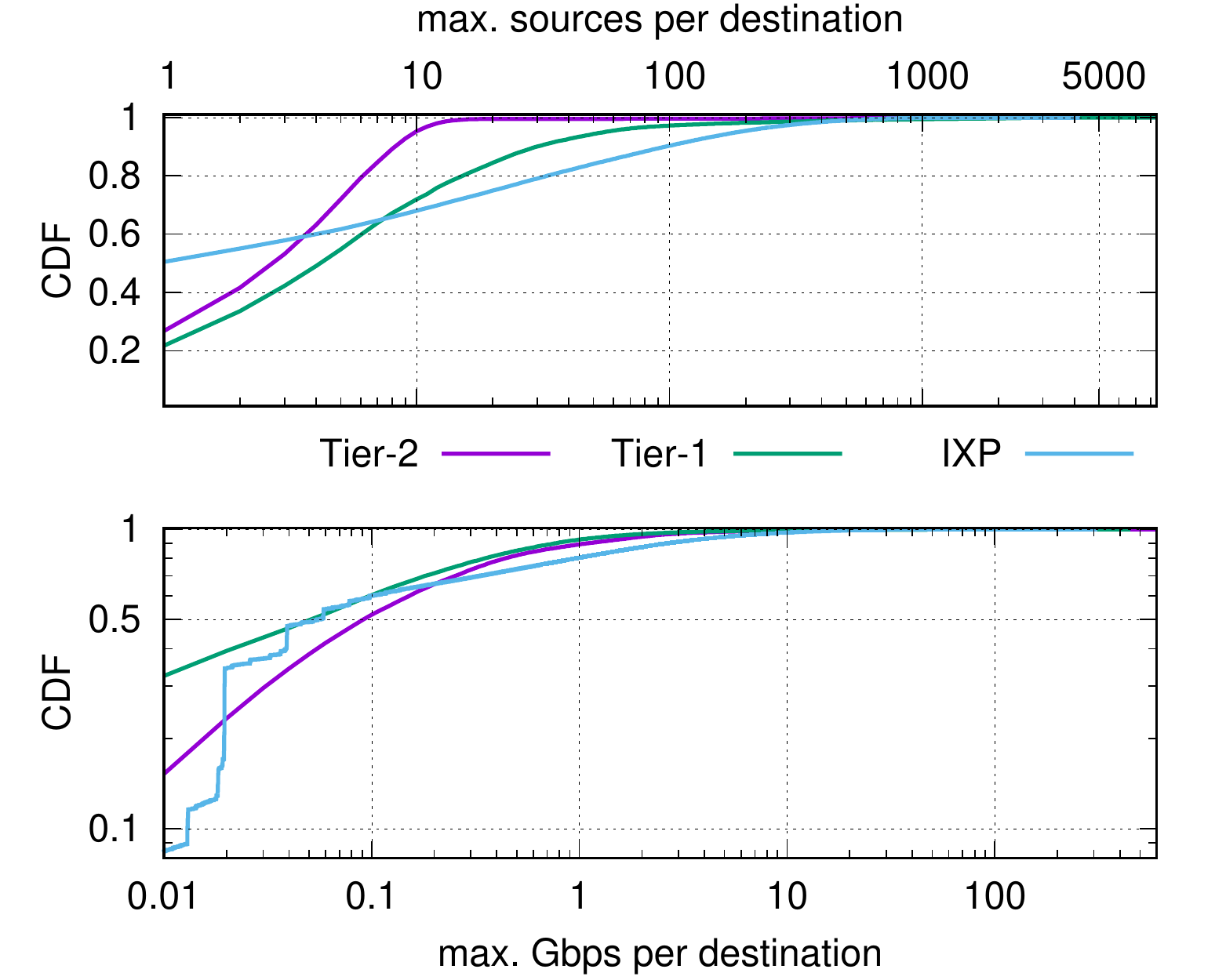}
    \label{fig:AttackSeverity_CDF}
  }

  \vspace{-1.5em}
  \caption{Classification of NTP amplification attacks at ISPs and an IXP.}
\vspace{-1em}
  \label{fig:eval}
}
\end{figure*}
The majority of NTP traffic ($80.81\%$) is delivered by our transit provider, the remaining $19.19\%$ are received over the IXP peering. Interestingly, $45.55\%$ of the peering traffic ($8.73\%$ over all) is coming from one AS while the median share per peer is $0.22\%$. Through the observations of the Memcached attack, we notice a shift of more traffic being transfered via the IXP peering ($88.59\%$), while one member alone accounts for $33.58\%$ of the total attack traffic. We assume that at the time this specific IXP member was exploited for Memcached amplification attacks. 
The observed traffic rates are significantly larger than the non-VIP booter attacks, but never reach the claimed bandwidth nor the advertised multiplication factors. Indeed, we experience only roughly $25\%$ of the traffic rate compared to that which was advertised.

\afblock{Amplification overlap.} Next we study how the different booters are related and to what extent they share the same amplifiers. 
We focus on NTP amplification attacks and compare the sets of reflectors used for the attacks. 
Figure~\ref{fig:refl_hmap} depicts $16$ independent self-attacks and the overlap of the NTP reflectors sorted by date.
We find stable set of reflectors for Booter $B$ with moderate churn of around $30\%$ over a time frame of two weeks (1), 
which suddenly uses a new set of reflectors (for booter $B$ from 18-06-12 to 18-06-13).
We also observe a churning set of reflectors over a long period (2).
The part marked by (3) shows same day measurements with a high overlap, which indicates a stability over short time frames and allows us to conclude that this booter doesn't randomly select reflectors from a larger set of reflectors, but uses the same reflectors for attacks in this time frame.
Moreover reflectors occasionally overlap between booter services (4).
Interestingly, the VIP and non-VIP services use the same set of reflectors. 
The difference in attack traffic is generated by a higher packet rate ($5.3$M pps compared to $2.2$M pps).
Moreover, the number of the reflectors used by booters is fairly low (in total $868$) compared to the globally available set of potential amplifiers (e.g., 9M potentially usable NTP servers according to \url{shodan.io} on May 12 2019). 
We conclude that identifying booter services according to their reflectors is difficult because reflectors are rotating quickly, are overlapping between different services and suddenly start using a new set of reflectors. Ultimately, this makes it impossible to identify specific booter traffic at a later point in time by using the set of reflectors we learn from the self-attacks.

\takeaway{Booters enable anyone to launch Gbps-scale DDoS for few dollars only.
We measure higher attack volumes than previously reported ($1.4$\,Gbps)~\cite{santanna2015booters}, even with non VIP-services.
We are the first to report the capabilities of a VIP booter services that peak at about $20$\,Gbps.
These traffic volumes are sufficient to take down web services and significantly disturb the operation of inter-domain links and Internet infrastructure.
We further find NTP-based amplification attacks to provide the most potent and reliable type of booter attacks. 
We believe this is because NTP amplifiers are more widespread and stable, while Memcached amplifiers focus on fewer networks, which in turn detect abuse more quickly and mitigate the problem.}
\section{DDoS Amplification Traffic}\label{sec:ixp_ddos}

Next, we study DDoS amplification traffic at our vantage points, i.e., a Tier-1 \& Tier-2 ISP, and an IXP. We focus our discussion on NTP traffic only, since most reliable booter-spawned attacks were executed over NTP. 
This analysis highlights the prevalence of attack traffic at our vantage points 
and continues to define filtering criteria before we study the FBI takedown in the next section.

\afblock{Optimistic NTP DDoS classification.}
In our self-attacks we observed amplified NTP packets to have a size of either $486$ or $490$ bytes (98.62\% over all observed packets) due to monlist requests. To put this in context of a realistic NTP traffic mix, we show the distribution of NTP packet sizes at the IXP in Figure~\ref{fig:NTP_packet_size_cdf}. We observe an almost bimodal distribution of $54\%$ of the NTP packets that are smaller than $200$ bytes (likely benign traffic) and $46\%$ that are larger (likely attack traffic). Thus, we define a threshold of $200$ bytes as an optimistic classification criterion.
\afblock{NTP amplification traffic in the wild.}
By applying our classification, we find $311K$ destinations (Tier-1 ISP: $36K$, Tier-2 ISP: $95K$, IXP: $244K$) that receive NTP reflection traffic for the measurement time (see Section~\ref{sec:data}).
We plot the destination IPs (victims) of the attack traffic in Figure~\ref{fig:ISP_AttackSeverity_scatterplor}.
The plot shows the number of unique amplification sources (y-axis) and the max traffic level in Gbps over one minute (x-axis). 
The majority of traffic peaks (y-axis) are between $5$ and $100$ Gbps (avg. $2.64$ Gbps).
Notably, $224$ victims receives more than $100$ Gbps, $5$ more than $300$ Gbps of attack traffic---a single destination even up to $602$ Gbps.
Note that only observing IXP (peering) traffic will underestimate the true attack volumes if (not captured) transit links are used (see Section~\ref{sec:self-attack}).

Higher traffic rates often involve a higher number of amplifiers (sources):
most destinations receive traffic from hundreds of amplifiers, some even from thousands (avg. $35$).
Outliers, particularly for the Tier-1 ISP, receive traffic from up to $\sim$$8500$ amplifiers. 

We next study the max number of amplifiers per attack target (unique dest. IP) within one minute bins, shown as CDF in Figure~\ref{fig:AttackSeverity_CDF} (top).
Most targets receive traffic from less than $5$ reflectors.
For the Tier-1 and the IXP about $70\%$ receive traffic from less than $10$.
Interestingly, for the Tier-2, $90\%$ of the targets receive amplification NTP traffic from less than 10 amplifiers.
For comparison, we depict the CDF of attack traffic peaks within one minute per target for all $3$ data sets in Figure~\ref{fig:AttackSeverity_CDF} (bottom).
Only a fraction of $0.09$ receives more than $1$ Gbps peak traffic and at the IXP we see $158$ targets receiving more than $100$ Gbps. On the other hand, a majority of targets receive a negligible amount of traffic.

\afblock{Conservative NTP DDoS classification.}
We aim at identifying a sample of NTP DDoS attacks with a low rate of false positives at the cost of false negatives.
Thus, we take a more conservative approach by excluding two more cases of false positives, the scanning or monitoring of NTP monlists~\cite{czyz2014} and the use of the NTP port by custom applications.
Based on insights from our self-attacks in Section~\ref{sec:booter}, we introduce two more filtering rules: (a) traffic to the target has to be larger than 1 Gbps and (b) has to originate from more than 10 amplifiers, where applying both rules reduces the number of NTP destinations by 78\% ((a) only: 74\%, (b) only: 59\%). This results in a data set likely only containing NTP DDoS traffic.
%\enlargethispage{2\baselineskip}

%
%

%
%

%
%

\takeaway{NTP-based DDoS traffic is prevalent at all three vantage points and can be classified using the proposed criteria. The observed attacks involve traffic rates of up to $600$ Gbps---often generated by a large number of reflectors---during our observation period. The extent and size of DDoS attacks we observe raises the question of whether the takedown of booters has an effect on the attack traffic in general.}
\section{FBI Takedown of 15 Booters}\label{sec:takedown}
On December 19, 2018 the FBI seized the domains of 15 booter websites (e.g., critical-boot.com or quantumstress.net)~\cite{usdoj2018misc}.
All booters seized were tested by the FBI prior to the takedown and ``the FBI determined that these types of services can and have caused disruptions of networks at all levels''~\cite{usdoj2018misc}.
Smaller seizures of single booter domains had occurred previously, e.g., the seizure of webstresser.org with more than 138k registered users by investigators in the U.S., U.K.\ and the Netherlands in 2018~\cite{webstresser}.
Beyond domain seizures, booter users and operators can face legal actions~\cite{legalActionsBooterUsers}, e.g., the operator of Titanium Stresser was sentenced to years in prison in 2017~\cite{titaniumStresser}.
Next we focus on studying the effects of seizures on the overall DDoS attack traffic.
That is, do booter take downs result in a significant reduction in DDoS attack traffic?
\subsection{Domain Perspective on Takedown}

\begin{figure}
	\includegraphics[width=0.85\columnwidth]{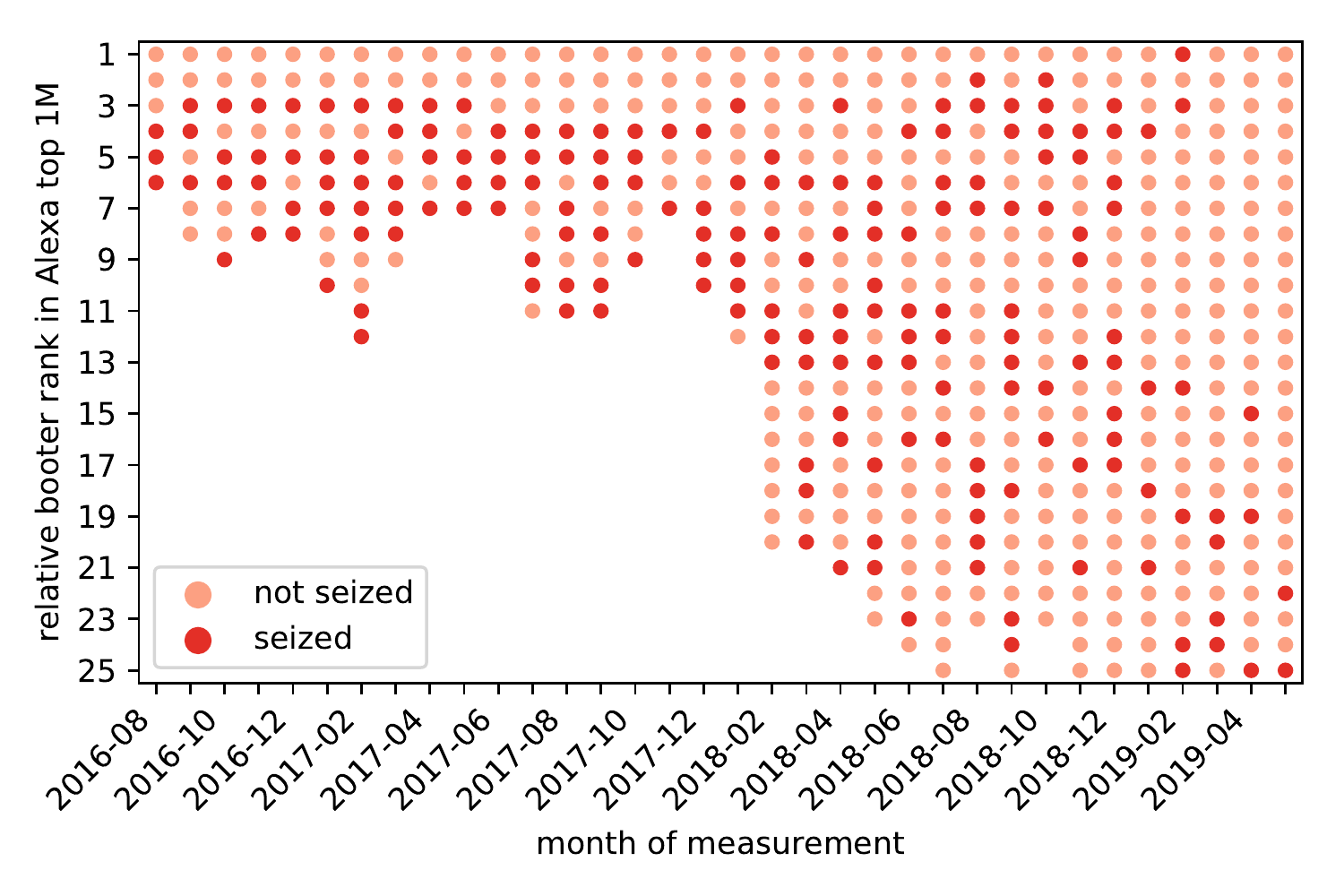}
	\vspace{-1.5em}
	\caption{Booter domains in the Alexa Top 1M by rank.}
	\vspace{-2em}
	\label{fig:domainrank}
\end{figure}

First, we take a control plane perspective on the available booter domains. We use weekly snapshots of all .com/.net/.org domains to identify booter websites by keyword matching following~\cite{booterBlacklist} (e.g., ``booter'', ``stresser'', ``ddos-as-a-service'').
This gives us an overview of booter domains before and after the takedown.
We identified 58 booter .com/.net/.org domains by manually visiting and verifying each domain matching the keyword search.
Using daily snapshots of the Alexa Top 1M list~\cite{toplists}, we rank the identified booter domains by their median Alexa rank over each month, shown in Figure~\ref{fig:domainrank}.
Booter domains that were seized in December are highlighted.
We observe the booter domains in the Alexa Top 1M to grow over time.
Seized domains have a high Alexa rank but not the highest among all booter domains---notably they occasionally still appear in the top 1M list (likely as a result of press reports pointing to those domains).
Thus, despite the seizure of 15\,domains, many alternative booter sites exist.
Following this, we select booter domains matching our keyword search after the takedown.
In this way, we identified a new domain for the seized booter A that became active after the takedown and entered the global Alexa Top 1M list on December 22---just three days after the seizure of their old domain.
The new domain was registered in June 2018 but remained unused until the takedown.
Our account credentials registered with the seized domain still work with the new domain (at the time of this writing).

\subsection{Traffic Perspective on Takedown}

For studying any data plane effects of the recent mass-seizure in 2018 at our vantage points, we do a time series analysis of 122 days beginning at Sep. 30, 2018 and ending at Jan. 30, 2019, spanning the seizure of the domains on Dec. 19, 2018. We calculate the following metrics:
\begin{inparaenum}[(a)]
	\item $wt_{30}$/$wt_{40}$ is a boolean metric indicating whether a one-tailed Welsh unequal variances test comparing the daily sum of packets 30/40 days before and 30/40 days after the takedown finds any significant difference at $p=0.05$;
	\item $red_{30}$/$red_{40}$ is the ratio of the daily average of sums of packets 30/40 days before and 30/40 days after the takedown.
\end{inparaenum}
Using these metrics, we investigate any combination of suspicious protocol ports (NTP, memcached, DNS, etc.) as source or destination port (to or from reflectors) for ingress and egress traffic. 
We start with a discussion of traffic \emph{to} DDoS reflectors, as we found significant changes for this type of traffic.

\begin{figure}[t!]
{
	\subfigure{
		\includegraphics[width=0.85\linewidth]{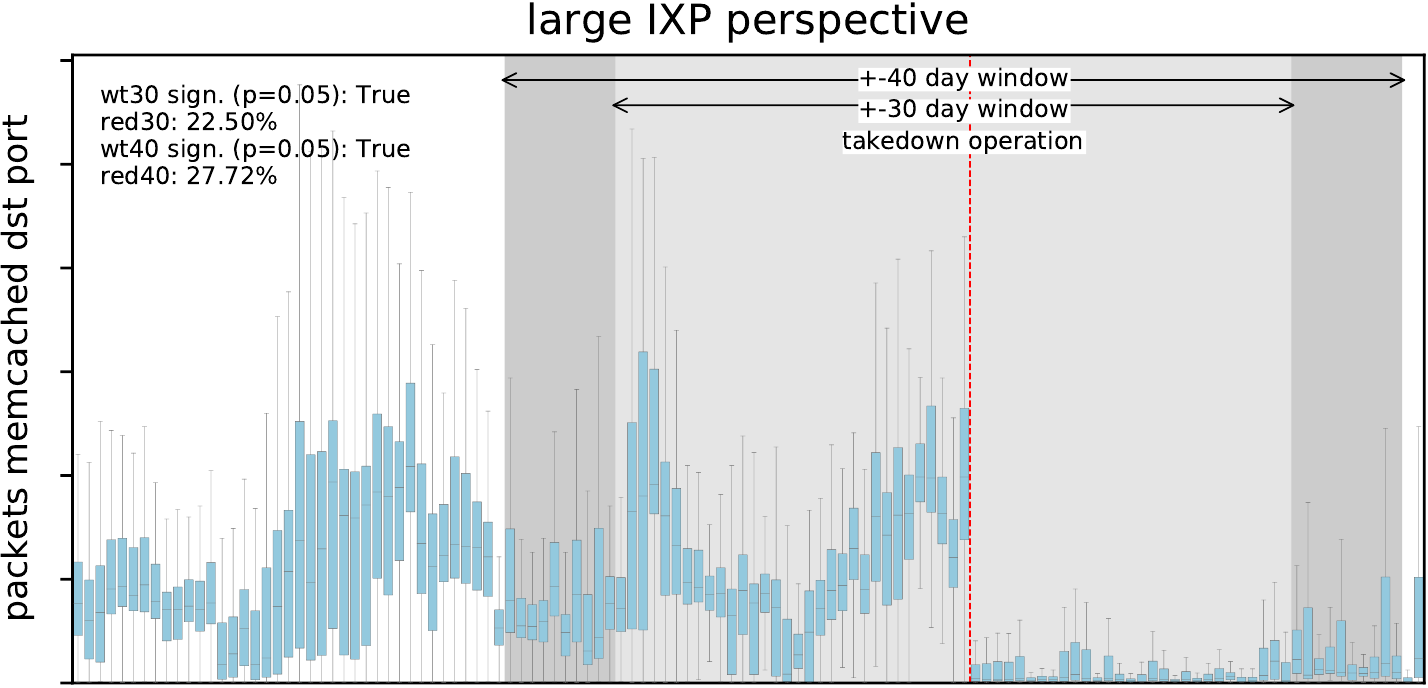}
		\label{fig:ixp_mem}
	}
	\subfigure{
		\includegraphics[width=0.85\linewidth]{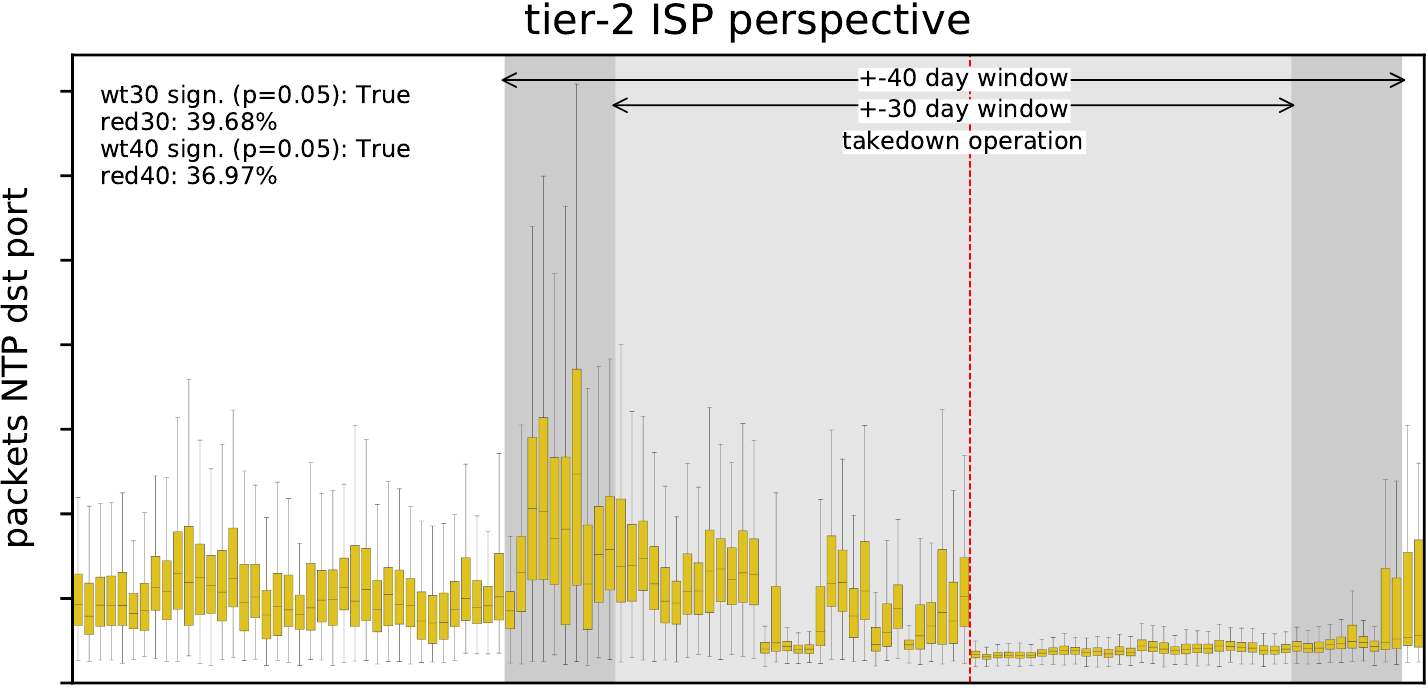}
		\label{fig:isp_ntp}
	}
	\subfigure{
		\includegraphics[width=0.85\linewidth]{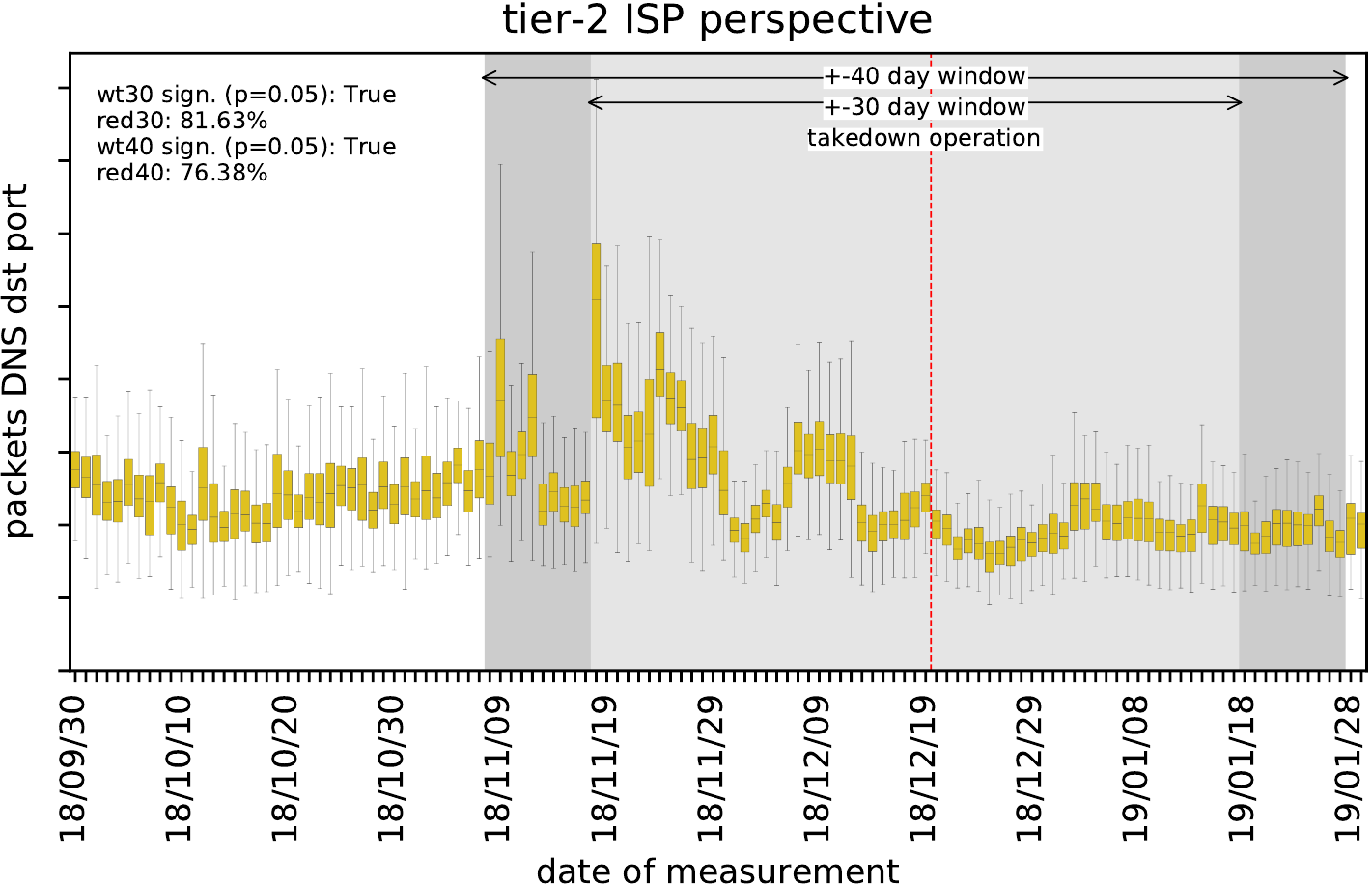}
		\label{fig:isp_ntp}
	}
	\vspace{-0.5cm}
	\caption{Selected significant changes in traffic before and after the takedown; $wt30$/$wt40$: significant lower packet counts at $p=0.05$ when comparing 30/40 days before and after the takedown; $red30/red40$: ratio of daily mean 30/40 days before and 30/40 days after the takedown.}
	\vspace{-0.5cm}
	\label{fig:eval}
}
\end{figure}

\afblock{Memcached traffic to reflectors.}
Memcached remains a popular attack vector due to its unsurpassed amplification factor.
As Memcached is a AS-internal object caching daemon, it is not expected to appear in regular inter-domain traffic.
Consequently, we assume that any UDP traffic with Memcached target port number 11211 is traffic flowing to a DDoS memcached reflector and accept possible noise added by scanning or other applications using the port.
We investigate the number of packets to memcached reflectors for the IXP vantage point in Figure \ref{fig:eval} (top).
A statistically significant reduction can be found for the 30 day window ($wt_{30}$) as well as for the 40 day window ($wt_{40}$).
The average daily number of packets after the takedown is 22.50\% ($red_{30}$) and 27.72\% ($red_{40}$) compared to before.
A comparable and significant reduction was found for the tier-2 ISP as well ($wt_{30}/wt_{40}=$True, $red_{30}=7.34\%$, $red_{40}=4.99\%$).

\afblock{NTP traffic to reflectors.}
NTP is a leading amplification vector due to the high number of open NTP reflectors.
We compare traffic to NTP reflectors under the assumption that any traffic with UDP target port 123 is a spoofed packet for triggering an attack, i.e., including false negatives such as legitimate NTP requests.
Even when accepting this unquantifiable amount of noise, we find significant reductions in traffic to NTP reflectors for the ISP vantage point (see \ref{fig:eval}, middle).
After the seizure, the number of packets falls to 39.68\% ($red_30$) and 26.97\% ($red_40$) respectively.
The same is true for the IXP vantage point ($wt_{30}/wt_{40}=$True, $red_{30}=22.5\%$, $red_{40}=27.72\%$).

\afblock{DNS traffic to reflectors.}
Similiar to NTP and Memcached, DNS is abused for DDoS as a reflector.
However, separating the comparably large share of legitimate from illegitimate requests is difficult.
Thus, our results for DNS request packets at the Tier-2 ISP's vantage point (Fig. \ref{fig:eval}, bottom) are not as visually impressive as for other vectors.
Nevertheless, we find a statistically significant reduction in traffic levels in both time windows ($wt_{30}$/$wt_{40}$).
In the week after the seizure, DNS requests fall to a global minimum. The overall reduction is larger than 20\% ($red_{30}/red_{40}$).
No reduction could be found for the IXP vantage point.

\begin{figure}[t!]
	\includegraphics[width=0.9\linewidth]{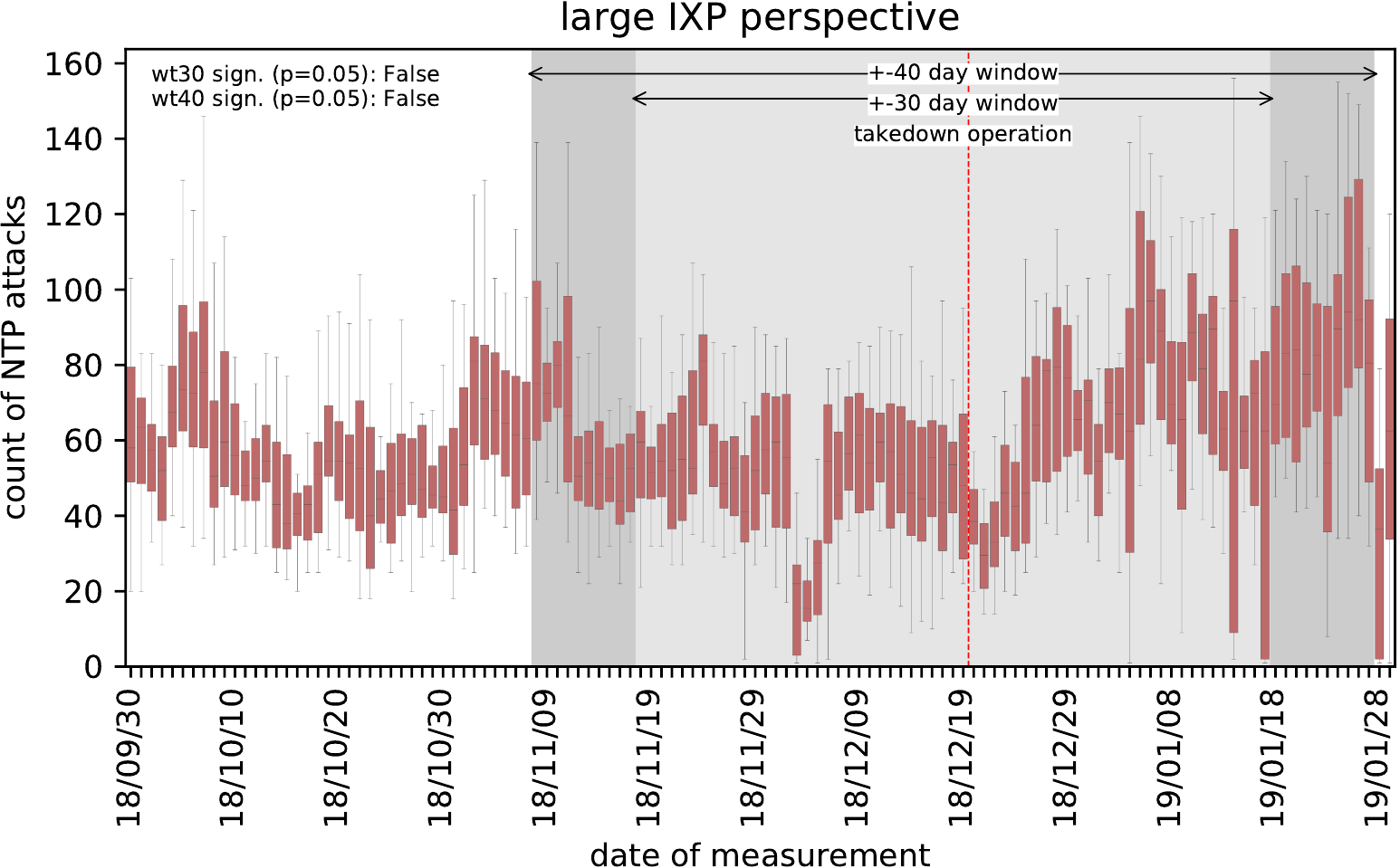}
	\vspace{-0.3cm}
	\caption{Systems under NTP DDoS attack per hour.}
	\label{fig:ixp_ntp_ddos_attacks_10refl_1G}
	\vspace{-0.5cm}
\end{figure}

\afblock{No significant reduction in attack volumes or number of systems attacked.}
A surprising finding of this work is that we find significant reductions for traffic flowing \emph{to} DDoS reflectors, but no significant reduction in attack traffic \emph{from} reflectors to victims, or in the number of systems attacked.
In order to minimize the probability of false conclusions with respect to this finding, we use the knowledge of NTP DDoS traffic characteristics from Section \ref{sec:ixp_ddos} to compose a filter for the number of systems under attack (see Fig. \ref{fig:ixp_ntp_ddos_attacks_10refl_1G}).
We isolate all IPs receiving NTP traffic with packets $>200$ bytes packet size from more than $\leq10$ hosts with more than $1$ Gbps traffic peak.
We do not find a significant reduction in the number of systems attacked ($wt30$/$wt40$).

\takeaway{The traffic patterns observed show a correlation with the FBI seizure.
We find significant reductions in DDoS traffic to possible DNS, NTP, and Memcached reflectors around the takedown operation.
Nevertheless, we could not find any significant reduction in traffic from reflectors to victims.
To exclude false positives, we use more reliable filters for NTP DDoS learned from our self-attacks, which shows no significant reduction after the takedown.
We conclude that seizing the front-end of Booter services does not improve the situation for DDoS victims, as the underlying infrastructure of reflectors remains online and can be utilized by third-parties without disruption.}

\section{Conclusions}\label{sec:concl}
This paper studies for the first time the effect of booter-based DDoS attacks through the lens of a major IXP, a tier-1 ISP, and a tier-2 ISP---with a focus on the effects of an FBI takedown of 15 booter websites in Dec. 2018.
By purchasing attacks
 against our own infrastructure from $4$ popular booters,
 we study booter capabilities.
The attack traffic levels generated by cheaper non-VIP services are considerably higher than reported in related work (avg. 1.4 Gbps)~\cite{santanna2015booters}.
We are the first to report the capabilities of a premium (VIP) booter service that peaks at 20\,Gbps while promising 60-80\,Gbps.
In our data sets, we observe NTP-based DDoS attack traffic to be prevalent at all three vantage points.
The attacks observed involve substantial traffic rates of up to 600\,Gbps during our observation period. 
To study if booter takedowns of law enforcement agencies help to reduce the attack traffic, we analyze the effect of an FBI-led mass-seizure of 15 booter domains in Dec. 2018 on NTP, DNS, and Memcached-based DDoS attacks.
We reveal that the takedown immediately had an effect on the DDoS amplification traffic especially reflectors.
However, it did not have any significant effect on DDoS traffic hitting victims or on the number of attacks observed. 
This shows that only seizing the front end is not enough as the underlying infrastructure of reflectors remains online and is utilized by third parties.
Moreover, we found at least one booter to become active under a new domain shortly after the seizure, while the number of booter service domains in total increased over the measurement period despite the seizure.
Our study aims to inform network operators to better understand the current threat-level, but also law enforcement agencies to recognize the need of additional efforts to shut down or block open reflectors.
Since our study is limited to technical parameters, the question arises whether this is sufficient to assess the health of the booter ecosystem. This motivates the need to better study the effects of law enforcement on the booter economy, e.g., on infrastructures, financing, or involved entities.
\section*{Acknowledgments}
We thank the anonymous reviewers of IMC and our shepherd, J. Alex Halderman (University of Michigan), for
their constructive comments, as well as our colleague Tim Tr\"aris, and our significant others for their tremendous patience.
This work was supported in part by the German Federal Ministry of Education and Research project AIDOS (6KIS0975K, 16KIS0976), by EC H2020 project CONCORDIA (GA 830927) and the SIDN fonds project DDoSDB (174058).
\bibliographystyle{acm}
\bibliography{references}

\end{document}